\documentclass[10pt,twocolumn,prl,aps,superscriptaddress,nofootinbib]{revtex4-1}
\usepackage{graphicx,amsmath,dcolumn}

\begin{document}

\newcommand{\eop}{{E_{0^+}}}
\newcommand{\dsdo}{\frac{d\sigma}{d\Omega}}

\title{Accurate Test of Chiral Dynamics in the \boldmath$\vec{\gamma} p
\rightarrow \pi^0p$ Reaction}

\author{D.~Hornidge}
\email{dhornidge@mta.ca}
\affiliation{Mount Allison University, Sackville, New Brunswick, E4L 1E6, Canada}
\author{P.~Aguar Bartolom\'e}
\affiliation{Institut f\"ur Kernphysik, University of Mainz, D-55128 Mainz, Germany}
\author{J.~R.~M.~Annand}
\affiliation{Department of Physics and Astronomy, University of Glasgow, Glasgow G12 8QQ, United Kingdom}
\author{H.~J.~Arends}
\affiliation{Institut f\"ur Kernphysik, University of Mainz, D-55128 Mainz, Germany}
\author{R.~Beck}
\affiliation{Helmholtz-Institut f\"ur Strahlen- und Kernphysik, University of Bonn, D-53115 Bonn, Germany}
\author{V.~Bekrenev}
\affiliation{Petersburg Nuclear Physics Institute, RU-188300 Gatchina, Russia}
\author{H.~Bergh\"auser}
\affiliation{II Physikalisches Institut, University of Giessen, D-35392 Giessen, Germany}
\author{A.~M.~Bernstein}
\affiliation{Department of Physics and Laboratory for Nuclear Science, Massachusetts
Institute of Technology, Cambridge, Massachusetts 02139, USA}
\author{A.~Braghieri}
\affiliation{INFN Sezione di Pavia, I-27100 Pavia, Italy}
\author{W.~J.~Briscoe}
\affiliation{The George Washington University, Washington, D.C. 20052-0001, USA}
\author{S.~Cherepnya}
\affiliation{Lebedev Physical Institute, RU-119991 Moscow, Russia}
\author{M.~Dieterle}
\affiliation{Institut f\"ur Physik, University of Basel, CH-4056 Basel, Switzerland}
\author{E.~J.~Downie}
\affiliation{Institut f\"ur Kernphysik, University of Mainz, D-55128 Mainz, Germany}
\author{P.~Drexler}
\affiliation{II Physikalisches Institut, University of Giessen, D-35392 Giessen, Germany}
\author{C.~Fern{\'a}ndez-Ram{\'i}rez}
\affiliation{Grupo de F{\'i}sica Nuclear, Universidad Complutense de Madrid, CEI Moncloa, E-28040 Madrid, Spain}
\author{L.~V.~Fil’kov}
\affiliation{Lebedev Physical Institute, RU-119991 Moscow, Russia}
\author{D.~I.~Glazier}
\author{P.~Hall Barrientos}
\affiliation{School of Physics, University of Edinburgh, Edinburgh EH9 3JZ, United Kingdom}
\author{E.~Heid}
\author{M.~Hilt}
\affiliation{Institut f\"ur Kernphysik, University of Mainz, D-55128 Mainz, Germany}
\author{I.~Jaegle}
\affiliation{Institut f\"ur Physik, University of Basel, CH-4056 Basel, Switzerland}
\author{O.~Jahn}
\affiliation{Institut f\"ur Kernphysik, University of Mainz, D-55128 Mainz, Germany}
\author{T.~C.~Jude}
\affiliation{School of Physics, University of Edinburgh, Edinburgh EH9 3JZ, United Kingdom}
\author{V.~L.~Kashevarov}
\affiliation{Lebedev Physical Institute, RU-119991 Moscow, Russia}
\affiliation{Institut f\"ur Kernphysik, University of Mainz, D-55128 Mainz, Germany}
\author{I.~Keshelashvili}
\affiliation{Institut f\"ur Physik, University of Basel, CH-4056 Basel, Switzerland}
\author{R.~Kondratiev}
\affiliation{Institute for Nuclear Research, RU-125047 Moscow, Russia}
\author{M.~Korolija}
\affiliation{Rudjer Boskovic Institute, HR-10000 Zagreb, Croatia}
\author{A.~Koulbardis}
\affiliation{Petersburg Nuclear Physics Institute, Gatchina, Russia}
\author{D.~Krambrich}
\affiliation{Institut f\"ur Kernphysik, University of Mainz, D-55128 Mainz, Germany}
\author{S.~Kruglov}
\affiliation{Petersburg Nuclear Physics Institute, Gatchina, Russia}
\author{B.~Krusche}
\affiliation{Institut f\"ur Physik, University of Basel, CH-4056 Basel, Switzerland}
\author{A.~T.~Laffoley}
\affiliation{Mount Allison University, Sackville, NB, E4L 1E6, Canada}
\author{V.~Lisin}
\affiliation{Institute for Nuclear Research, RU-125047 Moscow, Russia}
\author{K.~Livingston}
\author{I.~J.~D.~MacGregor}
\author{J.~Mancell}
\affiliation{Department of Physics and Astronomy, University of Glasgow, Glasgow G12 8QQ, United Kingdom}
\author{D.~M.~Manley}
\affiliation{Kent State University, Kent, Ohio 44242-0001, USA}
\author{E.~F.~McNicoll}
\affiliation{Department of Physics and Astronomy, University of Glasgow, Glasgow G12 8QQ, United Kingdom}
\author{D.~Mekterovic}
\affiliation{Rudjer Boskovic Institute, HR-10000 Zagreb, Croatia}
\author{V.~Metag}
\affiliation{II Physikalisches Institut, University of Giessen, D-35392 Giessen, Germany}
\author{S.~Micanovic}
\affiliation{Rudjer Boskovic Institute, HR-10000 Zagreb, Croatia}
\author{D.~G.~Middleton}
\affiliation{Mount Allison University, Sackville, NB, E4L 1E6, Canada}
\affiliation{Institut f\"ur Kernphysik, University of Mainz, D-55128 Mainz, Germany}
\author{K.~W.~Moores}
\affiliation{Mount Allison University, Sackville, NB, E4L 1E6, Canada}
\author{A.~Mushkarenkov}
\affiliation{INFN Sezione di Pavia, I-27100 Pavia, Italy}
\author{B.~M.~K.~Nefkens}
\affiliation{University of California Los Angeles, Los Angeles, California 90095-1547, USA}
\author{M.~Oberle}
\affiliation{Institut f\"ur Physik, University of Basel, CH-4056 Basel, Switzerland}
\author{M.~Ostrick}
\author{P.~B.~Otte}
\author{B.~Oussena}
\affiliation{Institut f\"ur Kernphysik, University of Mainz, D-55128 Mainz, Germany}
\author{P.~Pedroni}
\affiliation{INFN Sezione di Pavia, I-27100 Pavia, Italy}
\author{F.~Pheron}
\affiliation{Institut f\"ur Physik, University of Basel, CH-4056 Basel, Switzerland}
\author{A.~Polonski}
\affiliation{Lebedev Physical Institute, RU-119991 Moscow, Russia}
\author{S.~Prakhov}
\affiliation{University of California Los Angeles, Los Angeles, California 90095-1547, USA}
\author{J.~Robinson}
\affiliation{Department of Physics and Astronomy, University of Glasgow, Glasgow G12 8QQ, United Kingdom}
\author{T.~Rostomyan}
\affiliation{Institut f\"ur Physik, University of Basel, CH-4056 Basel, Switzerland}
\author{S.~Scherer}
\author{S.~Schumann}
\affiliation{Institut f\"ur Kernphysik, University of Mainz, D-55128 Mainz, Germany}
\author{M.~H.~Sikora}
\affiliation{School of Physics, University of Edinburgh, Edinburgh EH9 3JZ, United Kingdom}
\author{A.~Starostin}
\affiliation{University of California Los Angeles, Los Angeles, California 90095-1547, USA}
\author{I.~Supek}
\affiliation{Rudjer Boskovic Institute, HR-10000 Zagreb, Croatia}
\author{M.~Thiel}
\affiliation{II Physikalisches Institut, University of Giessen, D-35392 Giessen, Germany}
\author{A.~Thomas}
\author{L.~Tiator}
\author{M.~Unverzagt}
\affiliation{Institut f\"ur Kernphysik, University of Mainz, D-55128 Mainz, Germany}
\author{D.~P.~Watts}
\affiliation{School of Physics, University of Edinburgh, Edinburgh EH9 3JZ, United Kingdom}
\author{D.~Werthm\"uller}
\author{L.~Witthauer}
\affiliation{Institut f\"ur Physik, University of Basel, CH-4056 Basel, Switzerland}

\collaboration{A2 and CB-TAPS Collaborations}

\date{\today}

\begin{abstract}
A precision measurement of the differential cross sections $d\sigma/d\Omega$
and the linearly polarized photon asymmetry $\Sigma \equiv \left(
d\sigma_\perp - d\sigma_\parallel \right) \slash \left( d\sigma_\perp +
d\sigma_\parallel \right)$ for the $\vec{\gamma} p \rightarrow \pi^0p$
reaction in the near-threshold region has been performed with a tagged photon
beam and almost $4\pi$ detector at the Mainz Microtron.  The Glasgow-Mainz
photon tagging facility along with the Crystal Ball/TAPS multiphoton detector
system and a cryogenic liquid hydrogen target were used.  These data allowed
for a precise determination of the energy dependence of the real parts of the
$S$- and all three $P$-wave amplitudes for the first time and provide the most
stringent test to date of the predictions of chiral perturbation theory and
its energy region of agreement with experiment.
%
%
\end{abstract}
%

\maketitle

Low-energy pion photoproduction experiments are of special interest because
the pion, the lightest hadron, is a Nambu-Goldstone boson that by its
existence represents a clear signature of spontaneous chiral symmetry breaking
in QCD~\cite{Donoghue94}.  The dynamic consequences are that the production
and elastic scattering of neutral pions at low energies are weak in the
$S$ wave and strong in the
$P$ wave~\cite{Donoghue94,Bernard07,Bernard08,Scherer12}, as is seen in the
$\gamma N \rightarrow \pi N$ reaction~\cite{Scherer92,Bernstein07}.  In
neutral-pion photoproduction the $S$-wave threshold amplitudes are small since
they vanish in the chiral limit ($m_u, m_d \rightarrow 0$); their small, but
nonvanishing values are consequences of explicit chiral
symmetry breaking.  In addition, they are isospin
violating~\cite{Weinberg77,Bernard07,Bernard08} since $m_u \neq
m_d$~\cite{PDG2012,Leutwyler96}.  The magnitudes of low-energy scattering and
production experiments are predicted by chiral perturbation theory (ChPT), an
effective field theory of QCD based on spontaneous chiral symmetry
breaking~\cite{Scherer92,Donoghue94,Bernard07,Scherer12,Bernard08}.  Our
efforts have been focused on accurate measurements of low-energy $\gamma N
\rightarrow \pi N$ reactions, including the sensitive spin observables that
allow a unique separation of the $S$ and $P$ waves, to perform tests of these
predictions.  As has been stressed~\cite{Bernstein06,*BernsteinCD12}, any
serious discrepancy between these calculations and experiment must be
carefully examined as a challenge of our theoretical understanding of
spontaneous and explicit chiral symmetry breaking in QCD.

We have conducted an investigation of the $\vec{\gamma}p \rightarrow
\pi^0p$ reaction with the twin goals of obtaining 1) the energy dependence of
the photon asymmetry $\Sigma$ for the first time and 2) the most accurate
measurement to date of the differential cross section from threshold through
the $\Delta$ region.  The energy dependence of $\Sigma$, in combination with
the cross-section data, allowed us to extract the real parts of all $P$-wave
as well as the $S$-wave multipoles as a function of photon energy in the
threshold region.  These data in turn also allowed the first test of how well
ChPT calculations agree with the data as a function of photon energy above
threshold.  There exists one previous measurement of the photon
asymmetry~\cite{Schmidt01a}, but due to poor statistics resulting from small
cross sections and limited detector acceptance ($\approx 10\%$ for $\pi^0$
detection)\, $\Sigma$ was integrated over the entire incident photon energy
range, leading to data at only the bremsstrahlung-weighted energy of $E_\gamma
= 159.5$~MeV\@.  Moreover, the contribution to the asymmetry from the target
walls---significant in the threshold region---was not properly taken into
account~\cite{Erratum12}.  With the present setup the azimuthal acceptance
was vastly superior \textit{and} symmetric, the degree of linear polarization
was higher, a rigorous empty-target subtraction has been done, and as a result
both the statistical and systematic uncertainties are much smaller for
$\Sigma$ as well as the cross sections.  The most accurate previous
measurement along with a list of earlier efforts can be found in
Ref.~\cite{Schmidt01a}.

The experiment that is the focus of this Letter was conducted at the Mainz
Microtron MAMI~\cite{Herminghaus76,Kaiser08} where linearly polarized photons,
produced via coherent bremsstrahlung in a 100-$\mu$m-thick diamond
radiator~\cite{Lohmann94,Livingston08}, were sent through a 4-mm-diameter Pb
collimator and impinged on a 10-cm-long liquid hydrogen (LH$_2$) target
located in the center of the Crystal Ball~\cite{Unverzagt09}.  The TAPS
spectrometer served as a forward wall~\cite{Novotny91,*Novotny96}, and the
LH$_2$ target was surrounded by a particle identification
detector~\cite{Watts05}, used to differentiate between charged and uncharged
particles.  The incident photons were tagged up to an energy of 800~MeV using
the postbremsstrahlung electrons detected by the Glasgow-Mainz
tagger~\cite{Anthony91,*Hall96,*McGeorge08}.  For the electron beam of 855~MeV
used in this experiment, the tagger channels had a width of about 2.4~MeV in
the $\pi^0$ threshold region.  The diamond radiator was positioned relative to
the electron beam such that the photons produced had a polarization in the
range 50\%--70\% between the $\pi^0$ threshold and
$\simeq200$~MeV~\cite{Lohmann94}.

Neutral pions produced in the LH$_2$ target were identified in the detector
system using their 2$\gamma$ decay and a kinematic-fitting technique described
in detail in Ref.~\cite{Prakhov09}.  Both two- and three-cluster events that
satisfied the hypothesis of the reaction $\gamma p \to \pi^0p \to \gamma\gamma
p$ with a probability of more than 2\% were accepted as candidates for this
reaction.  Background contamination of the event candidates was found to be
from two sources:  interactions of the bremsstrahlung photons in the target
material different from liquid hydrogen, and accidental coincidences between
the tagger hits and the trigger based on the detector signals.  The background
was subtracted from the signal directly by using two different data samples,
the first of which included only events with accidental coincidences, the
second taken with an empty-target cell.

Acceptance of the detector system was determined by Monte Carlo simulation of
$\gamma p \to \pi^0 p$ using an isotropic angular distribution.  All events
were propagated through a {\sc GEANT}3.21 simulation of the experimental
setup, folded with resolutions of the detectors and conditions of the
trigger.  Close to the reaction threshold, the production-angle acceptance was
found to be almost uniform with a detection efficiency about 80\%.

The systematic uncertainties in the absolute numbers of the differential cross
sections for the reaction $\gamma p \to \pi^0 p$ obtained in the analysis of
the data set were estimated to be not larger than 4\%.  Such a magnitude of
the systematic uncertainty is mostly determined by the calculation of the
photon-beam flux, the experimental detection efficiency, and the number of
protons in the LH$_2$ target.  The systematic uncertainties in the numbers for
the photon asymmetry are on the level of 5\%, where this value comes mostly
from the uncertainty in the determination of the degree of the linear
polarization of the incoming photons.

\begin{figure}[htb]
\centerline{\includegraphics[width=0.48\textwidth]{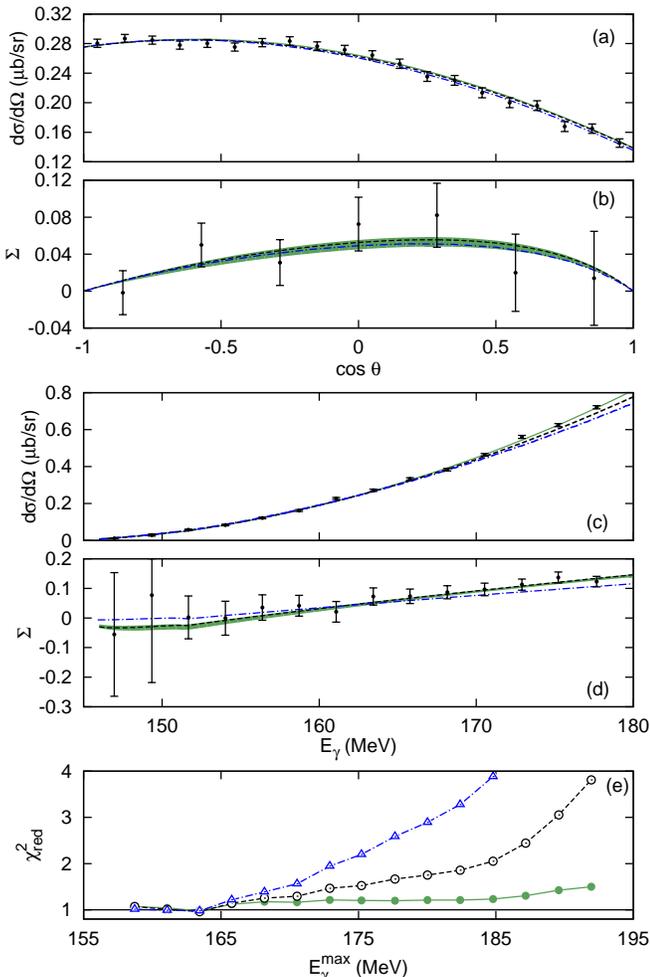}}
\caption{(color online) Differential cross sections in (a) $\mu$b/sr and (b)
photon asymmetries for $\pi^0$-production as a function of pion c.m.\
production angle $\theta$ for an incident photon energy of
$163.4\pm1.2$~MeV\@.  Energy dependence of the (c) differential cross sections
and (d) photon asymmetries at $\theta \simeq 90\pm3^\circ$.  Errors shown are
statistical only, without the systematic uncertainty of 4\% for
$d\sigma/d\Omega$ and 5\% for $\Sigma$.  The theory curves are dashed (black)
for HBChPT~\cite{Cesar-ChPT}, dash-dotted (blue) for relativistic
ChPT~\cite{Hilt,Hilt2}, and solid (green) for an empirical fit with an error
band.  (e) $\chi^2$ per degree of freedom for fits to the data in the range
from 150~MeV to $E_\gamma^\mathrm{max}$ for HBChPT~\cite{Cesar-ChPT} (open
black circles), relativistic ChPT~\cite{Hilt,Hilt2} (open blue triangles) and
an empirical fit (solid green dots), with lines drawn through the points to
guide the eye.  Note that in (c) and (d), the two points in incident photon
energy below the $\pi^+$ threshold are included; these two points are excluded
in the fits shown in (e) due to their large error bars.}
\label{fig:dxs_sigma}
\end{figure}

Results for the differential cross section and the photon asymmetry are
presented in Fig.~\ref{fig:dxs_sigma} as a function of the pion center-of-mass
(c.m.) production angle $\theta$ at $E_\gamma = 163.4\pm1.2$~MeV, and as a
function of incident photon energy at $\theta \simeq 90\pm3^\circ$.  Also
shown are one empirical and two theoretical fits: (1) heavy baryon chiral
perturbation theory (HBChPT) calculations to $O(q^4)$~\cite{BKM96,*BKM01} with
the five empirical low-energy constants brought up to date by fitting these
data~\cite{Cesar-ChPT}, (2) relativistic ChPT calculations [also to $O(q^4)$]
which as well have five low-energy constants fit to these
data~\cite{Hilt-thesis,Hilt2}, and (3) an empirical fit with error bands
calculated using the formalism from Refs.~\cite{Statistics12,Cowan}.  The
error bands take into account the correlations among parameters; details on
the method can be found in Appendix A of Ref.~\cite{Cesar-ChPT}.  Fits have
been performed employing a genetic algorithm combined with a gradient-based
routine that is thoroughly discussed in Ref.~\cite{Cesar-optimization}.

Because of the high quality of the present data, it is, for the first time,
possible to determine the energy range for which ChPT agrees with the data.
The values of the low-energy constants were obtained from fits to the data in
the range from 150~MeV to a variable $E_\gamma^\mathrm{max}$.  This was
done~\cite{Cesar-ChPT} using the $O(q^4$) formulas of heavy baryon
calculations~\cite{BKM96,*BKM01} and also for the relativistic
theory~\cite{Hilt,Hilt-thesis,Hilt2}.  Figure~\ref{fig:dxs_sigma}(e) displays
the $\chi^2$ per degree of freedom for the empirical fit and both ChPT
calculations.  For $E_\gamma^\mathrm{max}$ up to $\simeq$167~MeV, the ChPT
calculations are consistent with the empirical fit, but above this energy the
relativistic calculation starts to deviate from the data, whereas the heavy
baryon calculation begins to deviate at $\simeq$170~MeV\@.  This is
interesting, since the relative contributions of the terms containing the
low-energy constants of the relativistic calculation are significantly smaller
than those for the heavy baryon version, suggesting a better convergence for
the relativistic ChPT method.

The next step in the interpretation of the data is to extract the multipole
amplitudes and compare them to the theoretical calculations.  To set the
notation, the differential cross sections can be expressed in terms of the $S$-
and $P$-wave multipoles ($\eop,P_{1},P_{2},P_{3}$) and can be written as
\begin{equation}
\dsdo(\theta) = \frac{q}{k}(A+B\cos\theta+C\cos^2\theta),
\label{eq:dxs_abc}
\end{equation}
where $q$ and $k$ denote the c.m.\ momenta of the pion and the photon,
respectively.  The coefficients are given by $A=|\eop|^2+P_{23}^{\,2}$ with
$P_{23}^{\,2} = \frac{1}{2}\left(|P_2|^2 +|P_3|^2\right)$, $B=2\mathrm{Re}(\eop
P_1^\ast)$, and $C=|P_1|^2-P_{23}^{\,2}$.  The measurement of the cross
sections of earlier experiments~\cite{Schmidt01a} permitted the extraction of
$\eop$, $P_1$ and the combination $P_{23}$.  In order to extract the values of
Re$\eop$ and all three $P$ waves separately from the data, it is necessary to
also measure the photon asymmetry
\begin{equation}
\Sigma = \frac{d\sigma_{\perp}-d\sigma_{\parallel}}
	{d\sigma_{\perp} + d\sigma_{\parallel}}
 	= \frac{q}{2k}\left(|P_3|^2-|P_2|^2\right)
	\sin^2\theta/\dsdo(\theta),
\label{eq:asym}
\end{equation}
where $d\sigma_{\perp}$ and $d\sigma_{\parallel}$ are the differential cross
sections for photon polarization perpendicular and parallel to the reaction
plane with the pion and the outgoing proton.  To reiterate, the measurement of
the differential cross section \textit{and} $\Sigma$ allows for the separation
of the four multipoles.  It is important to note that the determination
reported here is more accurate than previous ones due to the far smaller
uncertainties of the cross sections as well as the energy dependence of
$\Sigma$.  Furthermore, we note that the $D$ waves have been neglected in both
(\ref{eq:dxs_abc}) and (\ref{eq:asym}), but they have recently been shown to
be important in the near-threshold region~\cite{Cesar-D,*Cesar-D2}.  Since
there are insufficient data to determine the $D$-wave multipoles empirically,
they have been taken into account by using their values in the Born
approximation, which is sufficiently accurate for the present analysis.

The empirical fits to the data employ the following ansatz for the $S$- and
$P$-wave multipoles
\begin{align}
\eop(W) &= E_{0^+}^{(0)} + E_{0^+}^{(1)}
	\left(
	\frac{E_\gamma-E_\gamma^\mathrm{thr}}{m_{\pi^+}}
	\right)
	+ i\beta\frac{q_{\pi^+}}{m_{\pi^+}}, \label{eq:swave} \\
P_{i}(W) &= \frac{q}{m_{\pi^+}} \left[P_{i}^{(0)}
	+ P_{i}^{(1)}\left(
	\frac{E_\gamma-E_\gamma^\mathrm{thr}}{m_{\pi^+}}
	\right)\right], \label{eq:pwaves}
\end{align}
where here $E_\gamma$ and $E_\gamma^\mathrm{thr}$ are in the lab frame, and
$E_{0^+}^{(0)}$, $E_{0^+}^{(1)}$, $P_{i}^{(0)}$, $P_{i}^{(1)}$ (with $i =
1,2,3$) are constants that are fit to the data.
[The empirical values are in units of $10^{-3}/m_{\pi^+}$:
$E_{0+}^{(0)}=-0.369 \pm 0.027$,
$E_{0+}^{(1)}= -1.47 \pm 0.13$,
$P_{1}^{(0)} =  9.806 \pm 0.068$, 
$P_{1}^{(1)} =  1.63 \pm 0.32$, 
$P_{2}^{(0)} = -10.673 \pm 0.070$, 
$P_{2}^{(1)} =-4.52\pm 0.31$,
$P_{3}^{(0)} = 9.671\pm0.060$, 
$P_{3}^{(1)} = 15.87\pm0.29$. 
The pairs $\left( E_{0+}^{(0)},E_{0+}^{(1)} \right)$ and $\left( P_{i}^{(0)},
P_{i}^{(1)}\right)$, $i =1,2,3$, are highly correlated.]

\begin{figure}[htb]
\begin{center}
\includegraphics[width=0.45\textwidth]{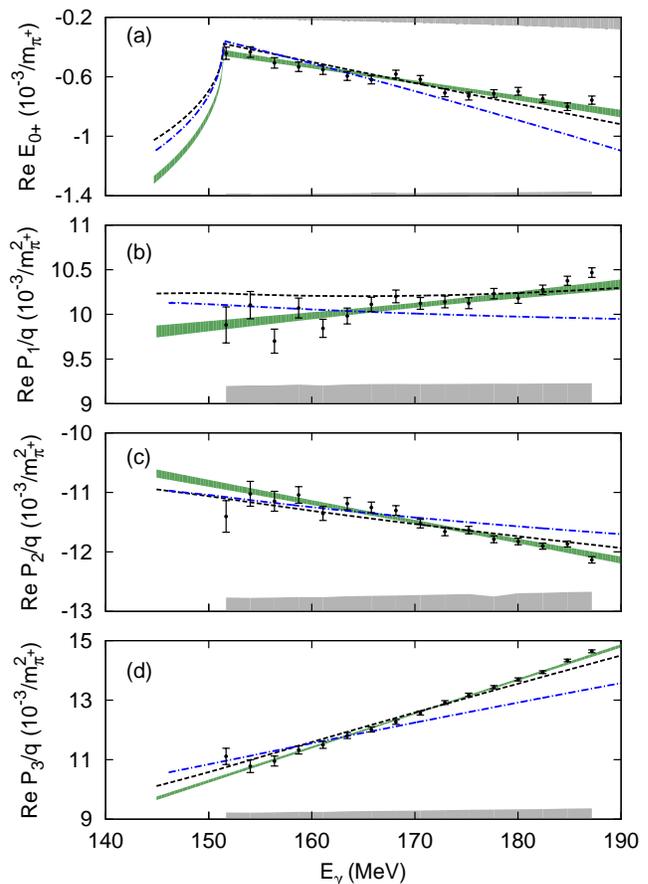}

\caption{(color online) Empirical multipoles as a function of incident photon
energy:  (a) Re$\eop$, (b) Re$P_1/q$, (c) Re$P_2/q$, (d) Re$P_3/q$.  The
points are single-energy fits to the real parts of the $S$- and $P$-wave
multipoles, and the empirical fits from Eqs.~(\ref{eq:swave}) and (\ref{eq:pwaves})
are shown with (green) statistical error bands.  The $\pm$ systematic
uncertainty for the single-energy extraction is represented as the gray area
above the energy axis, and the systematic uncertainty in the $S$-wave
extraction due to the uncertainty in the size of the $D$-wave contributions is
given by the gray area at the top of (a).  The theory curves are the same as
in Fig.~\ref{fig:dxs_sigma}.  Note that the two points in incident photon
energy below the $\pi^+$ threshold are excluded from all fits due to their
disproportionately large error bars.}

\label{fig:multipoles}
\end{center}
\end{figure}

Based on unitarity, the cusp parameter in Eq.~(\ref{eq:swave}) has the value
$\beta =  m_{\pi^+} a_\mathrm{cex}(\pi^+n \rightarrow \pi^0 p)
\text{Re} \eop(\gamma p \rightarrow \pi^+n) $~\cite{AB-lq}.  Using the
experimental value of $a_\mathrm{cex}(\pi^{-} p \to \pi^{0} n) =
-(0.122 \pm 0.002) /m_{\pi^+}$ obtained from the observed width in the $1s$
state of pionic hydrogen~\cite{Gotta}, assuming isospin is a good symmetry,
i.e.\ $a_\mathrm{cex}(\pi^{+} n \rightarrow \pi^{0} p) =
-a_\mathrm{cex}(\pi^{-} p \rightarrow \pi^{0} n)$, and the latest measurement
for $E_{0+}({\gamma} p \to \pi^{+} n) = \left( 28.06 \pm 0.27 \pm 0.45
\right)\times 10^{-3}/m_{\pi^+}$~\cite{Korkmaz}, we obtain $\beta = \left(
3.43 \pm 0.08 \right) \times 10^{-3}/m_{\pi^+}$, which was employed in the
empirical fit.  If isospin breaking is taken into
account~\cite{Hoferichter09, Baru11} we obtain $\beta=\left( 3.35 \pm 0.08
\right) \times 10^{-3}/m_{\pi^+}$.  In this experiment we do not have access
to the imaginary part of the $S$-wave amplitude and no difference
is found if either the isospin-conserving, the isospin-breaking, or even other
$\beta$ values such as those for dispersive effective chiral
theory, $\beta=3.10\times 10^{-3}/m_{\pi^+}$~\cite{LG} or HBChPT
$\beta=2.72\times 10^{-3}/m_{\pi^+}$, are employed.  Hence, the
uncertainty introduced by the errors in $\beta$ and isospin breaking is
smaller than the statistical uncertainties of the multipole extraction
depicted in Fig.~\ref{fig:multipoles}.

The extracted multipoles are displayed in Fig.~\ref{fig:multipoles} along with
the theoretical calculations.  The points are single-energy fits to the real
parts of the $S$- and $P$-wave multipoles, and the energy-dependent fits from
Eqs.~(\ref{eq:swave}) and (\ref{eq:pwaves}) are shown with the error band.
The imaginary part of the $S$-wave multipole $\eop$ was taken from unitarity
(\ref{eq:swave}) with the value of the cusp parameter explained above, the
imaginary parts of the $P$ waves were assumed to be negligible, and the
$D$-wave multipoles were calculated in the Born approximation.  The impact of
$D$ waves in the $P$-wave extraction is negligible~\cite{Cesar-D,*Cesar-D2}
but in the $S$ wave it can be sizeable.  In order to assess the uncertainties
in the $S$-wave extraction associated to our $D$-wave prescription, we have
estimated the uncertainty from the difference between the Born terms and the
Dubna-Mainz-Tapei dynamical model in Ref.~\cite{DMT2001}.  This error estimation is
depicted in Fig.~\ref{fig:multipoles} as a gray area at the top of the first
plot.  Note that the $D$ waves have a negligible impact in the $P$-wave
extraction (the uncertainty is smaller than the curve's width)
\cite{Cesar-D2}.

As was the case for the observables, there is very good agreement between the
two ChPT calculations and the empirical values of the multipoles for energies
up to $\simeq170$~MeV with the same pattern of deviations above that.

In conclusion, the combination of the photon asymmetry and improved accuracy
in the differential cross section has allowed us to extract the real parts of
the $S$-wave and all three $P$-wave multipoles as a function of photon energy
for the first time.  We have achieved an unprecedented accuracy in our
empirical extraction of the multipoles from the data, providing a more
sensitive test of the ChPT calculations than has previously been possible.
What we  have found is that none of the real parts of the multipoles $E_{0+}$,
$P_{1}$, $P_{2}$, $P_{3}$ is causing the gradual deviation from experiment
(increasing $\chi^{2}$) with increasing energy.  Rather, it is probably due to
the gradually increasing importance of the higher-order terms neglected in the
chiral series, or to the fact that the $\Delta$ degree of freedom is not being
taken into account in a dynamic way.

The authors wish to acknowledge the excellent support of the accelerator group
of MAMI, as well as many other scientists and technicians of the Institut
f\"ur Kernphysik in Mainz.  This work was supported by the Deutsche
Forschungsgemeinschaft (SFB~443), the Natural Science and Engineering Research
Council (NSERC) in Canada, the National Science Foundation and Department of
Energy in the United States, the U.K. Engineering and Physical Sciences
Research Council, the Schweizerischer Nationalfonds, the Spanish Ministry of
Economy and Competitiveness Grants No.\ JCI-2009-03910 and No.\
FIS2009-11621-C02-01, and the European Community Research Activity under the
FP7 Programme (Hadron Physics, Grant Agreement No.\ 227431).

\renewcommand{\bibname}{}
\bibliography{prl.bib}

\end{document}